# A Domain Specific Transformation Language

Bernhard Rumpe and Ingo Weisemöller

Software Engineering
RWTH Aachen University, Germany
http://www.se-rwth.de/

**Abstract.** Domain specific languages (DSLs) allow domain experts to model parts of the system under development in a problem-oriented notation that is well-known in the respective domain. The introduction of a DSL is often accompanied the desire to transform its instances. Although the modeling language is domain specific, the transformation language used to describe modifications, such as model evolution or refactoring operations, on the underlying model, usually is a rather domain independent language nowadays.

Most transformation languages use a generic notation of model patterns that is closely related to typed and attributed graphs or to object diagrams (the abstract syntax). A notation that reflects the transformed elements of the original DSL in its own concrete syntax would be strongly preferable, because it would be more comprehensible and easier to learn for domain experts. In this paper we present a transformation language that reuses the concrete syntax of a textual modeling language for hierarchical automata, which allows domain experts[1] to describe models as well as modifications of models in a convenient, yet precise manner. As an outlook, we illustrate a scenario where we generate transformation languages from existing textual languages.

## 1 Introduction and Problem Statement

Domain specific languages (DSLs) have the advantage of allowing domain experts to model parts of the system in a problem-oriented notation that is well-known in the respective domain. Like most documents in software development processes, models in DSLs underly frequent changes. These may include refactorings, automated modifications, or complex editing operations. Change operations on models can be described in explicitly defined model transformations.

To define a model transformation, we need an appropriate transformation language. Today's transformation languages [7] however operate on the abstract syntax and thus look very different from the DSL to be transformed. In the following sections, we are going to present an approach to close this gap.

---

[1] In our wording, the term "domain" refers to application domains such as business processes or a discipline of engineering as well as to technical domains such as relational databases or state based systems.

If the user wants to keep the look-and-feel of the DSL within the transformation language, then this language needs to embody elements of the concrete syntax of the underlying DSL, and is thus domain specific itself. Consequently, instead of having a single language for transformations of models in arbitrary DSLs, we would prefer a syntactically fitting transformation language that provides the same look-and-feel as the DSL at hand.

In this contribution, we state that the concrete syntax of a textual DSL can be reused to describe transformation rules, thus providing this look-and-feel. We substantiate our claim by the introduction of a transformation rule used in the process of flattening hierarchical automata and of the corresponding transformation language. Because the elements of the transformation language depend on the elements of the automata language in a systematic manner, we believe it is possible to systematically if not automatically derive the transformation language from a given DSL.

The following sections are outlined as follows: In Section 2 we provide a brief introduction to graph based model transformations, based on a rule used in the process of flattening hierarchical automata. We are going to reuse this example in the subsequent sections. Section 3 gives an introduction to existing approaches to the definition of model transformations in a domain specific notation. In Section 4 we explain what transformation rules in concrete syntax look like. In Section 5 we summarize the previous sections and give an overview of our ongoing and future work in this area.

## 2   Abstract and Concrete Syntax in Transformations

In the following, we consider *transformation rules* to be small steps of transformation in an appropriate language, which may be composed to more complex transformation sequences by control structures or rule application strategies. Composition mechanisms may vary (cf. [7, 17]), whereas we encounter some kind of transformation rules in almost any transformation language. Therefore and for reasons of space, we leave composition mechanisms out of consideration. Instead, we focus on the notation of transformation rules.

In graph based transformation approaches, rules consist of a left hand side (LHS) and a right hand side (RHS), which describe excerpts from a model that the transformation can be applied to (see [15, 12]). Informally explained, the LHS describes a part of the model before the application of the transformation rule, whereas the RHS describes the same part of the model after the rule application.

Because we basically describe excerpts from models, i.e. instances of a modeling language, in the LHS and RHS of a transformation rule, it seems natural to reuse the syntax of the modeling language when describing transformation rules. In current transformation approaches however, this reuse is limited to the abstract syntax for a variety of reasons, which means that the concrete syntax of the modeling language is not reflected in the transformation language.

We are going to show the difference between reusing the abstract syntax only and reusing both abstract and concrete syntax based on a transformation rule for hierarchical automata. The rule we consider is used in the process of flattening hierarchical automata, which is a simplified case of the transformations for flattening UML state machines (cf. [18, pp. 227 ff.] for details).

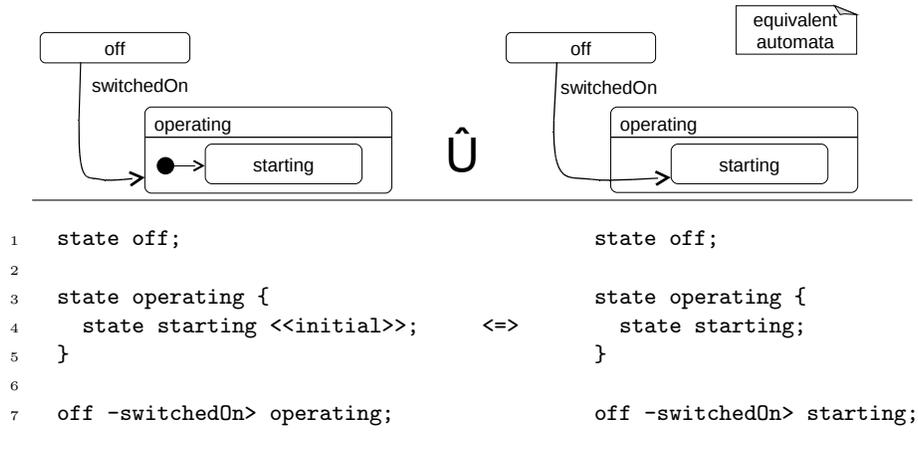

```
1   state off;                              state off;
2
3   state operating {                       state operating {
4     state starting <<initial>>;    <=>      state starting;
5   }                                       }
6
7   off -switchedOn> operating;             off -switchedOn> starting;
```

**Fig. 1.** Equivalent automata in graphical and textual representation

Before investigating the transformation rule itself, we take a look at the syntax of the DSL for hierarchical automata. Figure 1 shows both a graphical and textual representation of a hierarchical automaton on the left, and a graphical and textual representation of a semantically equivalent automaton on the right. The automaton on the right is obtained from the one on the left by forwarding the transition to the nested initial state. The upper part of the figure shows the automata in a graphical syntax, whereas the notation in the lower half is a textual representation of the same automata.

A model transformation rule that can transform an automaton on the left into the equivalent automaton on the right consists of two parts: a LHS, which matches a part of the automaton similar to the left side of Figure 1, and a RHS, which specifies the replacement, and which is similar to the right side of Figure 1. We ignore the RHS of the rule for the moment and take a look at the pattern matching part on the left only: Figure 2 shows the difference between a pattern based on the abstract syntax of the textual DSL from Figure 1, and the same pattern in a notation based on the concrete syntax of that language.

The language of the object diagram pattern in the upper part of the figure reuses the abstract syntax of the automata DSL. The same applies to the second notation (which we did not define explicitly, but is inspired by MOF QVT [11] and OCL [10]). Please note that these patterns are written in pseudocode rather than being executable by some tool, but they depict the general style of transformation languages based on the abstract syntax.

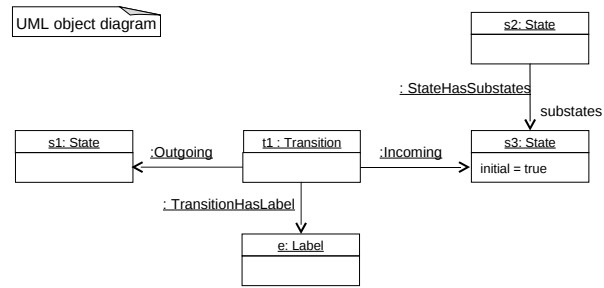

```
                  ─────── Automaton pattern in OCL-like abstract syntax ───────
1   s1 : State;
2   s2 : State;
3   s3 : State;
4   t : Transition;
5   e : Label;
6
7   s2.substates->contains(s3);
8   s3.initial = true;
9   t.source = s1;
10  t.target = s2;
11  t.label = e;
```

```
                  ─────── Automaton pattern in concrete syntax ───────
1   state $source;
2
3   state $outer {
4     state $inner <<initial>>;
5   }
6
7   $source -$event> $outer;
```

**Fig. 2.** Three variants of the LHS of a rule for transition forwarding

The statements in this pattern are either declarations of typed objects (ll. 1-5), links (l. 7) or additional constraints for these objects (ll. 8-11).

In comparison to this, the pattern based on the concrete syntax of the DSL, which is shown in the lower part of Figure 2, is more compact and easier to read. This is because the transformation language used here is close to the underlying modeling language rather than based on lists of objects, links and constraints.

The pattern matching language differs from the modeling language itself mainly in the use of *schema variables* such as $source that act as placeholders for concrete elements from the model.

## 3   Related Work

A number of publications addresses the specification of model transformations in a notation close to or identical to the corresponding modeling languages. This is usually referred to as *transformations in concrete syntax*, for instance by T. Baar and J. Whittle [3] as well as by M. Schmidt [19]. We will adopt this term for the remainder of this paper.

Both publications mentioned above adapt the concrete syntax of visual modeling languages for the specification of transformation rules. In either approach, the adaption of the syntax is performed manually. To our best knowledge, there is no implementation of either of these approaches available.

Several researchers have spent work on the derivation or inference of transformation rules from concrete examples [13, 22, 5]. In comparison to our work, these approaches usually require a manual adaption of the inferred rule in order to make it applicable to more models than the example it was derived from.

There are also approaches to define a specific transformation language manually, such as JTL [6] or the language for Java patterns presented in [2]. The manual definition of comparable languages for DSLs would be very tedious. Therefore, the generation of transformation languages that we present as an outlook in Section 5 can substantially save efforts for language developers.

Existing approaches to the generation of transformation languages that reflect the concrete syntax of the transformed models are currently limited to graphical, metamodel-based languages. The most mature approaches we are currently aware of are the ones by R. Grønmo [8] and by Kühne et al. [14]. These approaches however do not consider the concrete syntax of textual languages defined in grammars, and the generation of the languages is not fully automated.

Models in textual languages can also be transformed by term rewriting systems. E. Visser presents how term replacements can be written using the concrete syntax of the underlying language in [23]. Term rewriting rules however are usually limited to a connected (and typically small) subgraph of the target syntax tree, whereas in model transformations we often have to deal with rules that operate on objects distributed all over the syntax tree or even different input files.

Another example where the same language is used to describe expressions and transformations is mathematics and maybe proof systems close to mathematics [4, 16]. Mathematical equations can be understood as transformations, and indeed the success of mathematics to a large extent comes from a precisely defined, composable set of transformation rules (equalities) that allow to manipulate and simplify mathematical formulas in almost any form. Mathematics however does not need explicit references to the abstract syntax.

In conclusion, model transformations for textual languages could be substantially improved in terms of reflecting the concrete syntax of the underlying DSL, and — as far as can be seen from existing work — such transformation languages can to a wide extent be generated from the original modeling languages.

## 4 Syntactic Form of Transformation Rules

We now introduce the transformation rule language for hierarchical automata. This introduction is informal in the sense that it points out the style of transformation rules and what happens at execution time of these rules. We are not going to completely define their syntax and their semantics, but concentrate on the presentation style of transformation rules. For the understanding of this section, we assume that the reader has a basic knowledge of model transformations, especially model transformations based on graph transformations, and the application of transformation rules to host models as discussed in [1].

We pick up the example of forwarding transitions in automata to nested initial states (cf. Section 2). Figure 3 shows two possible notations of a transformation rule for the forwarding of a single transition, given in concrete syntax[2].

In our approach, which is shown in the upper half of Figure 3, a transformation rule consists of an integrated notation of its LHS and its RHS. In comparison to separate notations, as shown in the lower half of Figure 3, this has two major advantages: The first one is reduced redundancy between the LHS and the RHS, especially if we have a lot of elements that are not changed by the transformation and occur on both sides of the rule. The second one is the possibility to determine object identity between the LHS and the RHS: If an object does not have a name (such as the transition in lines 8 and 16 in the lower part of the figure), or if the name of an object is changed by the transformation rule, we have to introduce additional object IDs, such as $T in the example, for defining identical objects on the LHS and RHS.

In our example there are two differences between the LHS and the RHS. In the integrated notation, differences are indicated by a replacement inside the rule, denoted between square brackets [[...]] and the replacement operator :-. The first difference is defined in line 6: The state identified by $inner is not initial on the RHS, indicated by the removal of the modifier <<[[initial :- ]]>>. The second one occurs in line 9. The name of the target state of the transition modeled here is $outer on the LHS, but $inner on the RHS.

These differences describe exactly the modifications that are necessary to transform the LHS from our initial example (cf. Figure 1) into the RHS, i.e. into the automaton that has no nested initial states.

In contrast to our initial example from Figure 1, we do not have to use concrete identifiers of states or transition labels in the transformation rule. Instead of identifiers, we can use schema variables. In our transformation rule language, identifiers are interpreted as schema variables if and only if they start with a dollar sign. Thus, we can unambiguously distinguish between schema variables that must be matched when the transformation is executed, and fixed identifiers. Please note that schema variables cannot only be matched against identfiers, but against arbitrary syntactical elements. For example, $event in Figure 3 could also be matched against complex labels with events and preconditions.

---

[2] This is a simplified rule; actually a semantics preserving transformation would have to forward all incoming transitions to all nested initial states.

―――――――――――― Transformation rule in concrete syntax ――――――――――――
```
1 state $source;
2
3 state $outer {
4   state $inner << [[ initial :- ]] >>;
5 }
6
7 $source -$event> [[$outer :- $inner]];
```
―――――――――――――――――――――――――――――――――――――――――――――――――――――――――――――――

――― Transformation rule in concrete syntax, separated LHS and RHS ―――
```
1  match {
2    state $source;
3
4    state $outer {
5      state $inner << initial >>;
6    }
7
8    Transition $T [[ $source -$event> $outer; ]]
9  } replace {
10   state $source;
11
12   state $outer {
13     state $inner << initial >>;
14   }
15
16   Transition $T [[ $source -$event> $inner; ]]
17 }
```
―――――――――――――――――――――――――――――――――――――――――――――――――――――――――――――――

**Fig. 3.** Transformation rule for transition forwarding in concrete syntax with integrated notation of LHS and RHS (top) or separated notation (bottom)

When a transformation is executed, the transformation engine attempts to find a match for the pattern specified on the LHS. A schema variable may occur several times in a pattern, such as `$source` in lines 1 and 7. In this case, all matches against this schema variable in the host model must have the same value[3]. Elements that have a fixed identifier are matched against the element with exactly this identifier in the host model.

Once a match for the LHS is found, it is replaced by the RHS of the transformation rule. As our work does not focus on the definition of another transformation engine, but we need such an engine to demonstrate our approach, we have chosen a fairly standard way of interpreting the transformation as inspired by graph grammar tools like [21, 20, 1] in the first attempt.

---

[3] This part of the matching is currently limited to a flat, global namespace. We are going to to apply this concept to more complex namespaces in future work.

The graph matching approach allows (but not enforces) a match to have properties that are not given in the rule. For example, the initial state given in line 4 of the example may be mapped to a state that is both initial and final.

Our language also provides mechanisms to combine concrete syntax patterns with abstract syntax, thus allowing to define objects with abstract types or additional constraints referring to the abstract syntax. Moreover, it includes advanced concepts for pattern matching in attributed graphs such as sets of nodes or negative application conditions.

## 5 Conclusions, Current State and Future Work

In the previous section we gave an example of a transformation rule in concrete syntax, which is an instance of a domain specific transformation language. The systematic derivation of such transformation languages from DSLs as well as further improvements of our transformation engine are subject to our ongoing work in this area.

Our goal is to generate transformation languages from the grammars of DSLs. A configuration of the generation process (such as the specification of the variable prefix described above) would be acceptable, but the development of a transformation language in concrete syntax should not require writing source code in a programming language or modifying grammars manually.

We use the MontiCore tool set and framework for the language definition and all generation processes. MontiCore [9] allows for the integrated definition of the concrete and the abstract syntax of DSLs in a grammar format similar to EBNF. It also provides mechanisms to efficiently process models in these DSLs, for instance static analyses, code generation, or model transformations written in Java. Future versions of MontiCore will also provide support of model transformations as presented in this paper.

Figure 4 depicts the process of modeling and transformation language development as well as the usage of these languages according to our approach. A language developer defines the syntax of a DSL in a context-free grammar. From this definition, the rule language generator can automatically derive a grammar of a transformation language and a matching code generator, where the code generator also includes a language independent runtime environment. A domain expert can now not only define models in the DSL, but also implement model transformations in the generated language, which can be processed by the transformation language parser and code generator.

As a proof of concept we are currently working on more complex transformations, including the complete process of flattening UML state machines by transformations in concrete syntax. In order to develop more complex transformations in a manageable way, we are also working on a control flow language, which is syntactically and semantically close to a subset of Java and includes transformation rules as statements or expressions.

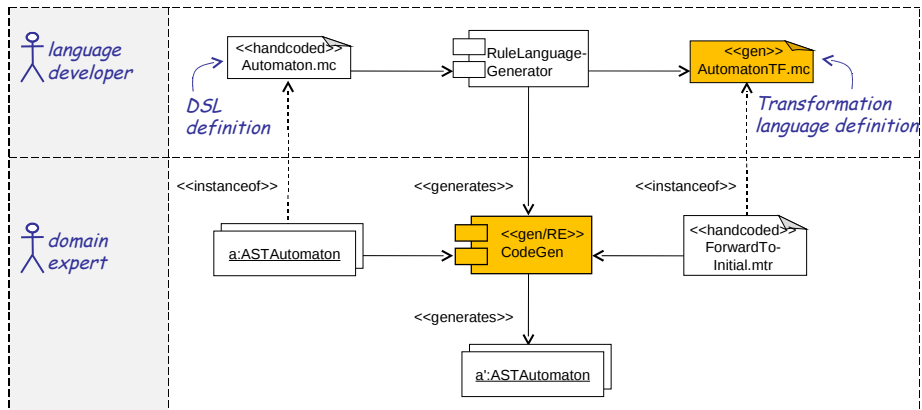

**Fig. 4.** Roles, documents and components in the transformation process

The current prototype can already execute a subset of the rules required to flatten UML state machines. We plan to come up with a more stable version that includes the control flow language and fully enables the transformations of state machines as well as some other languages in an upcoming version of MontiCore, which we plan to release by the end of this year.